\documentclass[usenatbib]{mn2e}
\usepackage{psfig}
\usepackage{amsmath}
\usepackage{amssymb}
\newcommand{\bm}[1]{\boldsymbol{#1}}
\newcommand{\pdrv}[2]{\frac{\partial #1}{\partial #2}}
\title[Simulations of molecular turbulence]%
{Hydrodynamical simulations of the decay of high-speed
molecular turbulence. I. Dense molecular regions}
 \author[Georgi Pavlovski et al.]%
{Georgi Pavlovski,$^1$%
\thanks{email: gbp@star.arm.ac.uk; mds@star.arm.ac.uk;
\mbox{mordecai@amnh.org};  rar@star.arm.ac.uk}
Michael D. Smith,$^1$
Mordecai-Mark Mac Low,$^2$
\newauthor   \& Alexander Rosen$^1$\\
$^1$ Armagh Observatory, College Hill, Armagh BT61 9DG,
Northern Ireland, U.K. \\
$^2$ American Museum of Natural History, Department of Astrophysics, 
79th St. at Central Park West, New York, NY 10024-5192, USA}
\date{Accepted .... . Received .... ; in original form .... }
\begin{document}
\maketitle
\begin{abstract}
We present the results from three dimensional hydrodynamical
simulations of decaying high-speed turbulence in dense molecular
clouds. We  compare our results, which include a detailed cooling 
function, molecular hydrogen chemistry and a limited C and O
chemistry, to those previously obtained for decaying isothermal 
turbulence.

After an initial phase of shock formation, 
power-law decay regimes are uncovered, as in the isothermal case.
We find that the turbulence decays faster than in the isothermal 
case because the average Mach number remains higher, due to the 
radiative cooling. The total thermal energy, initially raised by 
the introduction of turbulence, decays only a little slower
than the kinetic energy. 

We discover that molecule reformation, as the fast turbulence
decays, is several times faster than that predicted for a non-turbulent
medium. This is caused by
moderate speed shocks which sweep through a large fraction of the
volume, compressing the gas and dust. 
Through reformation, the molecular density
and molecular column appear as complex patterns of filaments, clumps and
some diffuse structure. In contrast, the molecular fraction has a wider
distribution of highly distorted clumps and copious
diffuse structure, so that density and
molecular density are almost identically distributed
during the reformation phase. 
We conclude that molecules form in swept-up clumps
but effectively mix throughout via subsequent expansions 
and compressions.

\end{abstract}
%
%
%-----------------------------------------------------------------
\section{Introduction}\label{sec:intro}
An understanding of turbulence is of fundamental importance 
to many research areas in astrophysics. In particular, compressible
turbulence plays a central role in the process by which stars are
formed (reviewed by \cite{VazSem00, Padoan01}). 
Stars form out of dense clouds of molecular gas, which have formed
out of diffuse clouds of atomic gas. The turbulent energy deduced
from observations of the molecular gas is sufficient to delay the
gravitational collapse, making molecular turbulence of great
interest. The purpose of this study is to provide insight through
the first three-dimensional simulations of molecular
turbulence.

Isothermal or polytropic equations of state have been the central
premise of previous investigations of turbulent models for molecular
clouds (e.g. \cite{VazSem96, M-MML98, Padoan98, Stone98}).
This simplification has the advantage that parameter space can be 
explored with numerical simulations, and the influence of magnetic
fields and gravity can be determined in some detail  (e.g. 
\cite{Bals01, Passot95, Kless00a, Heitsch01} ). The isothermal
approximation is indeed valid in many specific models where
molecular cooling is efficient and low gas temperatures are
maintained. Provided the temperature is low, one can argue that
an isothermal equation of state is as good as any other, with
little feedback from the the thermal pressure on the dynamics.

However, for cloud formation and evolution molecular chemistry and 
cooling is critical \citep{Lang00, Lim01, Lim99}. 
Molecular hydrogen forms most
efficiently where the gas is warm but the grains are cool (H$_2$
forms mainly when atoms combine after colliding and sticking to
dust grains).  Simple molecules like OH, CO and H$_2$O form in the
gas phase with H$_2$ as the reactive agent. These molecules are not
only important coolants, but associated emission lines provide a
means of measuring the cloud properties. Molecules are dissociated
as a consequence of fast shocks, UV radiation, X-rays and cosmic
rays \citep{Herbst00}. We thus need to study molecular turbulence to
determine the distribution and abundances of molecular species. 

Even in cool optically thin regions, as assumed here,
molecular pressures may influence the dynamics. Strong
shock waves may be driven from hot
atomic regions, formed by shock waves, into the proximity of
cold molecular gas. Areas devoid of dust, or in which
molecule formation is inefficient (e.g. because atoms
will evaporate off warm grains rather than
recombine) may attain and maintain pressures as high as
turbulent pressures.

The rate of decay of supersonic turbulence is important to 
the theory of molecular clouds.
A possible consequence of the rapid decay of kinetic energy
is that the turbulent clouds we observe have short lives.
The short lives, however, may not provide sufficient time for
relaxation into thermodynamic and chemical equilibrium. Hence, 
cloud structure and content evolve simultaneously  as a cloud
evolves dynamically, rather than proceeding through a
series of equilibrium states. Therefore, conclusions based on
assuming a molecular fraction to be a function of density alone
(e.g. \cite{Balles99a}) might not always be accurate. We note,
however, that the conclusion reached by \cite{Balles99a} of 
rapid cloud formation is supported by this study.

Decaying supersonic turbulence is the subject of this initial study.
The work is a direct extension, in terms of code and initial
conditions, of  isothermal simulations presented by \cite{M-MML98}.
Our goals here are as follows:
\begin{itemize}

\item To directly compare the decay of the kinetic energy $E_{kin}$
with those derived in the papers of \citep{M-MML98} and
\citep*{Smith00I, Smith00II}.
\item To determine the distribution of the molecular column density
as opposed to the total column density \citep{VazSem01}
\item To study the rate of formation of molecules.
\end{itemize}
We omit in this study magnetic field, self-gravity and
photodissociating radiation. We begin with
a fully molecular cube of dense gas and 
apply an initial turbulent field of
velocity perturbations. The turbulence was chosen
to be sufficient to dissociate the gas in one case
(root mean square (rms) speed of 60\,km\,s$^{-1}$) and to leave
the gas molecular in another case (rms speed of 15\,km\,s$^{-1}$).

%-----------------------------------------------------------------
\section{Methods}
\subsection{The equations}
Numerically we solve the time-dependent flow equations:
\begin{align}
\label{equ:hydro1}
\pdrv{\rho}{t} + \nabla\cdot\left(\rho\,\bm{v}\right) &= 0, \\
\label{equ:hydro2}
\pdrv{\left(\rho\,\bm{v}\right)}{t} + 
      \left(\bm{v}\cdot\nabla\right)\bm{v}
&= -\frac{1}{\rho}\nabla p, \\
\label{equ:hydro3}
\pdrv{e}{t} + \bm{v}\cdot\nabla\,e &= -p\nabla\cdot\bm{v} 
+ \Lambda\left(T, n, f\right), \\
\label{equ:hydro4}
\pdrv{\left(f n\right)}{t} + \nabla\cdot\left(f n \bm{v}\right) &= 
R\left(T, n, f\right) - D\left(T, n, f\right),
\end{align}
where $n$ is the hydrogen nuclei density, $e$ is the internal
energy density and $f$ is the molecular hydrogen abundance (i.e.
$n_{\rmn{H}_2} = f n$). We consider the gas as a
mixture of atomic and molecular hydrogen with 10\% of helium
(i.e. $n_{\rmn{He}} = 0.1\,n$). Therefore, the total
particle density is $n_{\rmn{tot}} = (1.1 - f)\,n$, mass density
is $\rho = 1.4 n m_{\rmn{H}}$ and the
temperature is $T = p\,/\,(k\,n_{\rmn{tot}})$. An
internal energy loss term has been added 
to the r.h.s. of the energy flow
equation: $\Lambda$ is the loss of energy 
through radiation and chemistry
per unit volume. The function consists of 11 distinct parts, summarised
below. $R$ and $D$ are reformation and dissociation rates of
molecular hydrogen respectively.  A detailed
description of the chemistry can be found in \cite{Smith02}. Formulae
for $\Lambda$, $R$ and  $D$ are given in the Appendix~\ref{appx}
%
%-----------------------------------------------------------------
\subsection{The numerical model: ZEUS-3D}\label{ssec:zeus}
%-----------------------------------------------------------------
As a basis, we employ the ZEUS-3D code \citep{Stone92I, Stone92II}.
This is a second-order, Eulerian, finite-difference code.
We study here compressible hydrodynamics
without gravity, self-gravity or thermal conduction. No physical
viscosity is modelled, but numerical viscosity remains present,
and a von~Neumann artificial viscosity determines the dissipation
in the shock front.

Further coding for molecular chemistry and molecular and atomic cooling
has been added.  Our ultimate goal is to develop a reliable code
with which we can tackle three dimensional molecular dynamics,
later adding self-gravity, magnetic fields,  ambipolar diffusion and
radiation. We thus restrict the cooling and chemistry lists to
just those items essential to the dynamics. We have employed
the simultaneous implicit method  discussed by
\cite{Sutt97} in which the time step is adjusted so
as to limit the change in internal energy in any zone to 30\%.

The cooling is appropriate for dense cloud material of any
atomic-molecular hydrogen mixture. We include H$_2$
ro-vibrational and dissociative cooling, CO and H$_2$O
ro-vibrational cooling, gas-grain, thermal bremsstrahlung
and a steady-state approximation to atomic cooling
\citep[see][Appendix~A]{Smith02} and Appendix~\ref{appx} of
the article.

We take a very basic network of chemical reactions.
Time-dependent hydrogen chemistry is included, but C and O
chemistry is limited to the reactions with H and H$_2$
which generate OH, CO and H$_2$O \citep[see][Appendix~B]{Smith02} 
and Appendix~\ref{appx} of the article.
Equilibrium abundances are calculated, which are generally accurate
within the shocks where molecules are rapidly formed and destroyed.
In cold molecular gas, however, the available oxygen
will probably remain primarily in the form of water even if it
has not been shock-processed. Therefore, the code is constructed
to follow the shock-enhanced chemistry and cooling rather than the
cold molecular gas. This will generally not be a problem at the
high densities where CO cooling and gas-grain heating and cooling
determine the properties of the low-temperature gas.

The formula for H$_2$ formation is critical to the results.
Reformation  takes place on grain surfaces with the
rate taken from \citep{Holl79}:
\begin{multline}\label{eqn:kR}
k_R = \big(3 \times 10^{-18}\,[{\rm cm}^3 {\rm s}^{-1}]\,\big) \times \\
\frac{T^{0.5} f_a}{1 + 0.04\,(T+T_g)^{0.5}+ 2\times10^{-3}\,T +
8\times10^{-6}\,T^2}
\end{multline}
with a factor 
\begin{equation}
f_a = \big(1 + 10000\exp(-600/T_g)\big)^{-1},
\end{equation}
which means, with $T_g = 20$~K assumed, that $f_a$ is quite close
to unity  in our simulations. Recent experimental results, summarised by
\cite{Katz99}, indicate that the H$_2$ formation rate could be
considerably lower. The inherent grain mass and 
uniform space distribution
adopted here also influence the results. 
This would imply that the absolute reformation times
derived here may need revision; we intend to explore a
range of possibilities in a following study.

The computations were performed on a Cartesian grid with
uniform initial density and periodic boundary conditions
in every direction, to simulate a region internal to a molecular
cloud. The hydrogen was fixed to be 
{\em fully molecular} at the beginning 
(fraction: $f = 0.5$).
Initial stress was introduced by Gaussian perturbations
applied to model velocities with a flat spectrum extending over
a narrow range of wave numbers given by $ 3 \le |\bm{k}| \le 4$.
We have performed runs with root mean square velocities of
15 km~s$^{-1}$, 30 km~s$^{-1}$ and 60 km~s$^{-1}$, and 
computational domain sizes of 32$^3$, 64$^3$, 128$^3$ 
and 256$^3$ zones. The number density has been taken to be 
$n = 10^6$~cm$^{-3}$, physical box size $L = 10^{16}$~cm, and
the initial temperature has a homogeneous distribution of $T_i = 100$~K.
Abundances for helium, free oxygen and carbon
were taken as 0.1, $5\times10^{-4}$ and  $2\times10^{-4}$.
%-----------------------------------------------------------------
\section{Data analysis}
%-----------------------------------------------------------------
\subsection{Decay rates}\label{ssec:rates}
%
%===========================================
\begin{figure*}
\begin{minipage}[h]{0.75\linewidth}
\centering\psfig{file=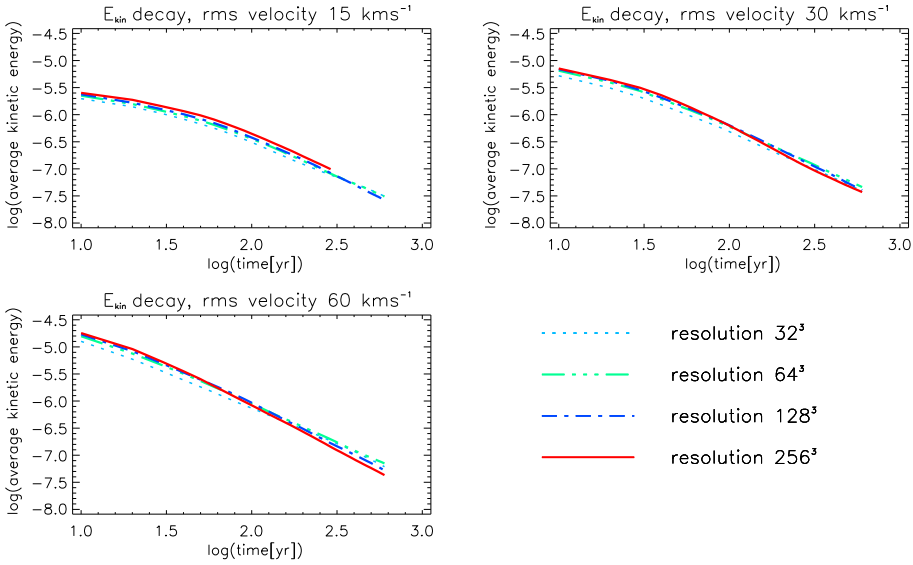}
\end{minipage}\hfill
\begin{minipage}[t]{0.25\linewidth}
\caption{Average kinetic energy as a function of time; log--log plot.}
\label{fig:kin} %Fig 1
\end{minipage}
\end{figure*}
%
%-------------------------------------------
\begin{table}
\caption{Power law indices $\eta$ of kinetic energy decay. 
The $\eta$ coefficients are derived from linear fitting 
on the time interval  $[90, 600]$ years. Average error $\pm0.006$.}
\label{tab:kin}
\begin{tabular}{cccc}
\multicolumn{1}{c}{resolution} & \multicolumn{3}{c}{initial rms
velocity for the run} \\
          & 15 km s$^{-1}$  & 30 km s$^{-1}$  & 60 km s$^{-1}$  \\
32$^3$    & $-1.19$         & $-1.33$         & $-1.33$ \\
64$^3$    & $-1.32$         & $-1.38$         & $-1.36$ \\
128$^3$   & $-1.45$         & $-1.48$         & $-1.52$ \\
256$^3$   & $-1.34^\star$   & $-1.51$         & $-1.60$ \\
\end{tabular}\\
$^\star$ -- linear fitting on time interval $[90, 300]$ years;
the run terminated half way due to numerical difficulties\\
\end{table}
%
%-------------------------------------------
%\begin{table}
%\caption{Power law indices $\eta$ of kinetic energy decay derived
%from curve fit to formular \ref{f:kin}. Average error $\pm0.005$.}
%
%\label{tab:kin2}
%
%\begin{tabular}{cccc}
%\multicolumn{1}{c}{resolution} & \multicolumn{3}{c}{initial rms
%velocity for the run} \\
%          & 15 km s$^{-1}$  & 30 km s$^{-1}$  & 60 km s$^{-1}$  \\
%32$^3$    & $-1.46$         & $-1.51$         & $-1.36$ \\
%64$^3$    & $-1.60$         & $-1.57$         & $-1.48$ \\
%128$^3$   & $-1.74$         & $-1.67$         & $-1.64$ \\
%256$^3$   & $-1.86$         & $-1.89$         & $-1.79$ \\
%\end{tabular}\\
%\end{table}
%===========================================
\begin{figure*}
\begin{minipage}[h]{0.75\linewidth}
\centering\psfig{file=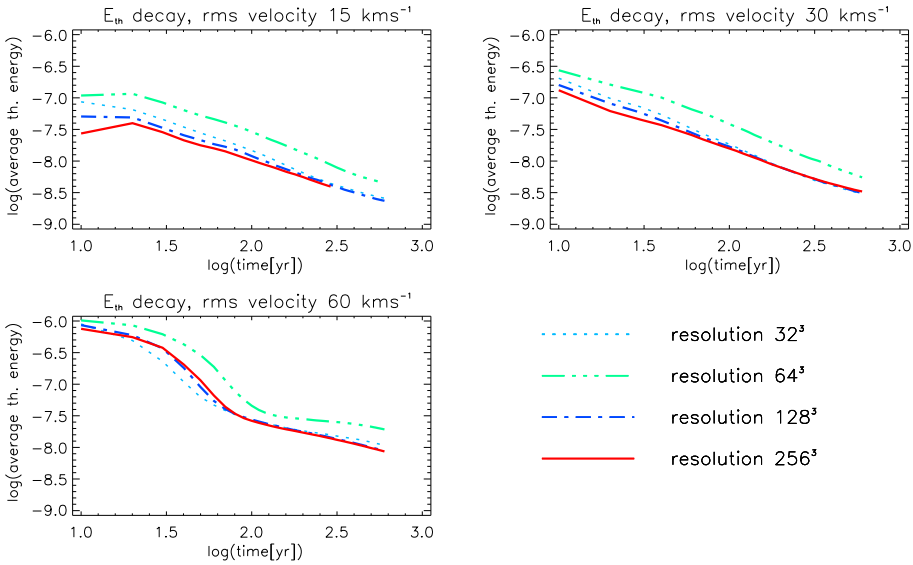}
\end{minipage}\hfill
\begin{minipage}[t]{0.25\linewidth}
\caption{Average thermal energy as a function of time; 
log--log plot.}
\label{fig:eth} %Fig 2
\end{minipage}
\end{figure*}
%
%-------------------------------------------
\begin{table}
\caption{Power law indices $\mu$ of thermal energy decay.
The $\mu$ coefficients are derived from linear fitting performed
on the time interval $[90, 600]$  years. Average error $\pm0.008$.}
\label{tab:eth}
\begin{tabular}{cccc}
\multicolumn{1}{c}{resolution} & \multicolumn{3}{c}{initial rms
velocity for the run} \\
          & 15 km s$^{-1}$  & 30 km s$^{-1}$  & 60 km s$^{-1}$  \\
32$^3$    & $-0.78$         &  $-0.82$        & $-0.47^\dagger$ \\
64$^3$    & $-0.97$         &  $-0.99$        & $-0.34^\dagger$ \\
128$^3$   & $-0.82$         &  $-0.81$        & $-0.63^\dagger$ \\
256$^3$   & $-0.85^\star$   &  $-0.75$        & $-0.62^\dagger$ \\
\end{tabular}\\
$^\star$ -- linear fitting on time interval $[90, 300]$ years;
the run terminated half way due to numerical difficulties\\
$^\dagger$ -- linear fitting on the time interval $[140, 600]$
years, present final decay behaviour after initial strong cooling
regime (see Fig.~\ref{fig:eth})
\end{table}
%
%============================================
\begin{figure*}
\begin{minipage}[h]{0.75\linewidth}
\centering\psfig{file=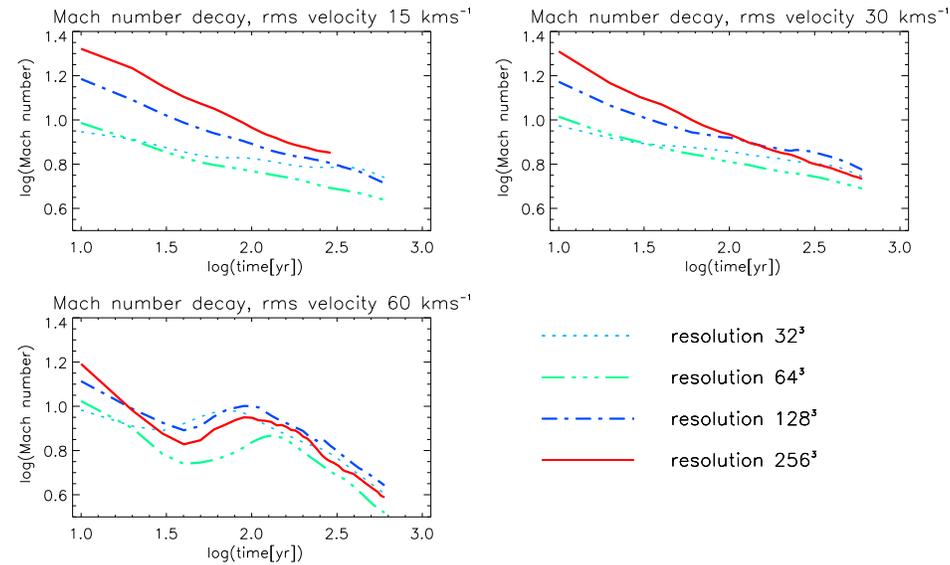}
\end{minipage}\hfill
\begin{minipage}[t]{0.25\linewidth}
\caption{Average Mach number as a function of time; log--log plot.}
\label{fig:mach} %Fig 3
\end{minipage}
\end{figure*}
%
%-------------------------------------------
\begin{table}
\caption{Power law indices $\nu$ of Mach number decay.
The $\nu$ coefficients are derived from linear fitting performed
on the time interval $[90, 600]$  years. Average error $\pm0.005$.}
\label{tab:mach}
\begin{tabular}{cccc}
\multicolumn{1}{c}{resolution} & \multicolumn{3}{c}{initial rms
velocity for the run} \\
          & 15 km s$^{-1}$  & 30 km s$^{-1}$  & 60 km s$^{-1}$  \\
32$^3$    & $-0.08$         & $-0.13$         & $-0.46^\dagger$ \\
64$^3$    & $-0.16$         & $-0.15$         & $-0.55^\dagger$ \\
128$^3$   & $-0.21$         & $-0.16$         & $-0.49^\dagger$ \\ 
256$^3$   & $-0.23^\star$   & $-0.24$         & $-0.54^\dagger$ \\
\end{tabular}\\
$^\star$ -- linear fitting on time interval $[90, 300]$ years;
the run terminated half way due to numerical difficulties\\
$^\dagger$ -- linear fitting on the time interval $[140, 600]$
years, present final decay behaviour after initial strong cooling
regime (see Fig~\ref{fig:mach})
\end{table}
%
%===========================================
\begin{figure}
\centering\psfig{file=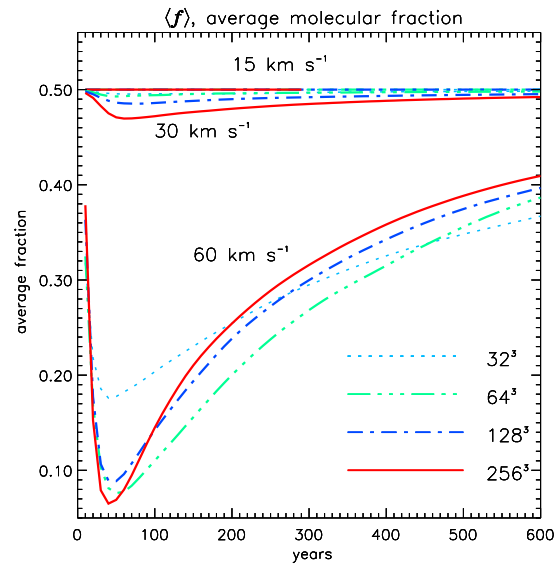}
\caption{Average molecular fraction as a function of time.
There is virtually no dissociation  during the run with an
initial rms of 15 km~s$^{-1}$ -- all four resolution curves
merge in one straight line to the 0.5 level. 
30 km~s$^{-1}$ -- intermediate case; 
60 km~s$^{-1}$ -- practically all molecules are temporarily destroyed.
}
\label{fig:frac} %Fig 4
\end{figure}
%
%===========================================
%
The decay in total kinetic energy is displayed in
Fig.~\ref{fig:kin}. An initial phase of evolution during which
waves steepen and velocity gradients transform
into shocks is apparent. The
shock formation time is short, with a timescale
\begin{equation} \label{f:tiphys}
t_i \sim \frac{L}{2\,k_m\,v_{rms}}
     = 45.3\times\frac{L}{10^{16}\,{\rm [\rmn{cm}]}}
        \frac{10\,[{\rm km\,s}^{-1}]}{v_{rms}}
        \frac{3.5}{k_m}\ [\rmn{yr}].
\end{equation}
where $k_m$ is defined as the initial mean wavenumber
\citep{M-MML98, Stone98, Smith00I}.
Hence, shocks develop faster
in the 60~km~s$^{-1}$ turbulent field, as evidenced
by the immediate power-law decay.

A rapid power-law decay of kinetic energy with time was uncovered
in the isothermal case \citep{Stone98, M-MML98, Smith00I}. This
actually followed an initial phase of evolution during which
velocity gradients increase and shocks appear. The time scale
for decay is of the order of a wave crossing time, which implied that
turbulent molecular clouds must be short lived if the 
turbulence is not regenerated.

The first result here is the confirmation that
molecular turbulence  behaves similar to isothermal turbulence,
but with higher rates of decay. The
curves of average {\it kinetic energy} decay are shown in 
Fig.~\ref{fig:kin}. Power laws deduced from linear fitting
($\log E_{kin}\propto \eta \log t$) of the
curves  are shown in Table~\ref{tab:kin}.

We have found that the formula
\begin{equation}\label{f:kin}
E_{kin} = E_0 ( 1 + \frac{t}{t_i}) ^ \eta ,
\end{equation}
fits the data quite well. However, values of $| \eta |$ 
derived from from fitting  the curve (\ref{f:kin}) are 
slightly larger due to nonzero values of $t_i$. 
This functional dependence explains why values of 
the average kinetic energy at the end are compatible, while at the 
beginning they are different by the factor of 4.

The power-law index $\eta$ increases with $v_{rms}$. This is
consistent with the high power-law index found
for high Mach number, $M = 50$, isothermal turbulence
\citep{Smith00I}.

The second result is that the total thermal energy ($E_{th} = 
nkT(3.3 - f)/2$) also decays
with time , as opposed to the imposed constant
value for the isothermal case.
The average  {\it thermal energy} density behaviour is
presented in Fig.~\ref{fig:eth} and the  corresponding decay rates
$\mu$: $E_{th}\propto t^{\mu}$ are shown in Table~\ref{tab:eth}.
The thermal energy aids our description of the physics.
We have found that it  also has a power-law decay
for the low speed case in which molecules are not
destroyed. The thermal energy decays, although not as fast as the
kinetic energy. Thus the Mach number remains
high relative to the isothermal case, which may well
account for the steeper kinetic energy decay rate. Eventually, however,
the gas cools down to just below the grain temperature
and the turbulent heating input falls off. At this temperature
range, however, the numerical code is not accurate enough to resolve the
minimum.

The fact that the thermal energy 
exhibits a dependence on  numerical resolution
(to a higher degree than kinetic energy does) 
is due to the importance of
cooling layer resolution for the energy balance and smoothening effects
on coarser grids. This implies lower values of thermal energy on finer
grids, although a smoothening effect for the 
$64^3$ run led to higher (than the
$32^3$ run) values of the average thermal energy.

With an initial rms speed of 60 km\,s$^{-1}$, the
molecules predominantly dissociate within 40 years. Cooling
associated with the dissociation and shock heating are competing
processes (Fig.~\ref{fig:eth}). After 40 years, there
is a steep  decay in thermal energy as trace CO and H$_2$O
molecules efficiently cool gas between 100~K and 8000~K.
As confirmation, we note that the number of zones within which
the temperature is very high ($>10^3$ K) falls from a percentage
$\approx$ 45\%  to $\approx$ 1\% during the first
150~yr.  The  gas which had become almost fully atomic (see plots
of average molecular  fraction in Fig.~\ref{fig:frac} and histograms
on  Fig.~\ref{fig:histmf}), is again over 60\% molecular hydrogen
after 300~yr. As the molecules reform,
the gas is heated  since the molecule energy released
on reformation is  channelled
collisionally into the gas (rather than being radiated) at the
high density of these simulations. Therefore, the thermal energy
subsequently decays slower than in all other cases, approaching
a shallow power-law value. These changes are not directly reflected
in the kinetic energy; however, the rapid gas cooling maintains
a high Mach number, which leads to the fast decay of kinetic
energy.

For reference, the energies are plotted in erg~cm$^{-3}$,
with the initial kinetic energy given by
\begin{multline}
E_i = \frac{1}{2} \rho v_{rms}^2
  = 1.2 \times 10^{-6}
  \left(\frac{n_o}{10^6\,[\rmn{cm}^{-3}]}\right)\times\\
    \left(\frac{v_{rms}}{{\rm 10\,[\rmn{km}\,s}^{-1}]}\right)^2
    [{\rm erg\,cm}^{-3}]
\end{multline}
and the final (thermal) energy, after complete molecule reformation
($\gamma = 10/7$), is
\begin{multline}
E_f = \frac{1}{1-\gamma} n\,k\,T
  = 1.6 \times 10^{-9}\frac{n_o}{10^6\,[\rmn{cm}^{-3}]}\times\\
  \frac{T_f}{10\,[\rmn{K}]}\;[{\rm erg\,cm}^{-3}].
\end{multline}
Plots of the average {\it Mach number} are shown in
Fig.~\ref{fig:mach}, 
corresponding decay rates, $\nu$: $M\!a\propto t^\nu$ are in 
Table~\ref{tab:mach}. Note that the display is logarithmic,
with an average  Mach number exceeding 10 maintained for 
the first 100\,yr in the low-speed case.
Here we define the average Mach number as
\begin{equation}\label{mach:equ}
M\!a = \sqrt{\frac{\langle \bm{v}^2 \rangle}
                  {\langle  c_s^2   \rangle}}\, ,
\end{equation}
where
\[
c_s^2 =\frac{(3.5 - 3f)(2.2 - 2f)}{(3.3 - 2f)^2}\frac{e}{\rho}
\]
is an adiabatic speed of sound, and the bracket notation 
$\langle\dots\rangle$ means averaging over the domain:
$L^{-3}\sum_{i,j,k}$.
The Mach number decaying behaviour can be qualitatively explained 
by noticing that it should be roughly proportional to the 
fraction $\left(E_{kin} /E_{th}\right)^{1/2}$, although 
the equality $\nu = (\eta - \mu) / 2$ is not generally true. 

The increase of the Mach number during the 
period $\approx[30, 120]$ yr in the runs
with rms velocity of 60 km~s$^{-1}$ is due to the
 strong dissociative cooling at that
time, as shown by the thermal energy decay curves, i.e. where 
$|\mu| > |\eta|$. Note that this case, with the
highest initial Mach number, possesses the lowest Mach number
once the shocks have formed. This is due to the dissociation,
generating a warm atomic gas. The question then is: why do the
turbulent motions decay so fast in this case, despite the low
Mach number? The suggestion is that the shocks are preferentially
running into the  denser molecular gas, which can  more efficiently
dissipate the energy because it is cooler. In isothermal gas, 
however, this effect would not be apparent.

%-----------------------------------------------------------------
\subsection{Molecular hydrogen evolution}\label{ssec:mh}
%
%-----------------------------------------------------------------
\begin{figure*}
\begin{raggedleft}
\psfig{file=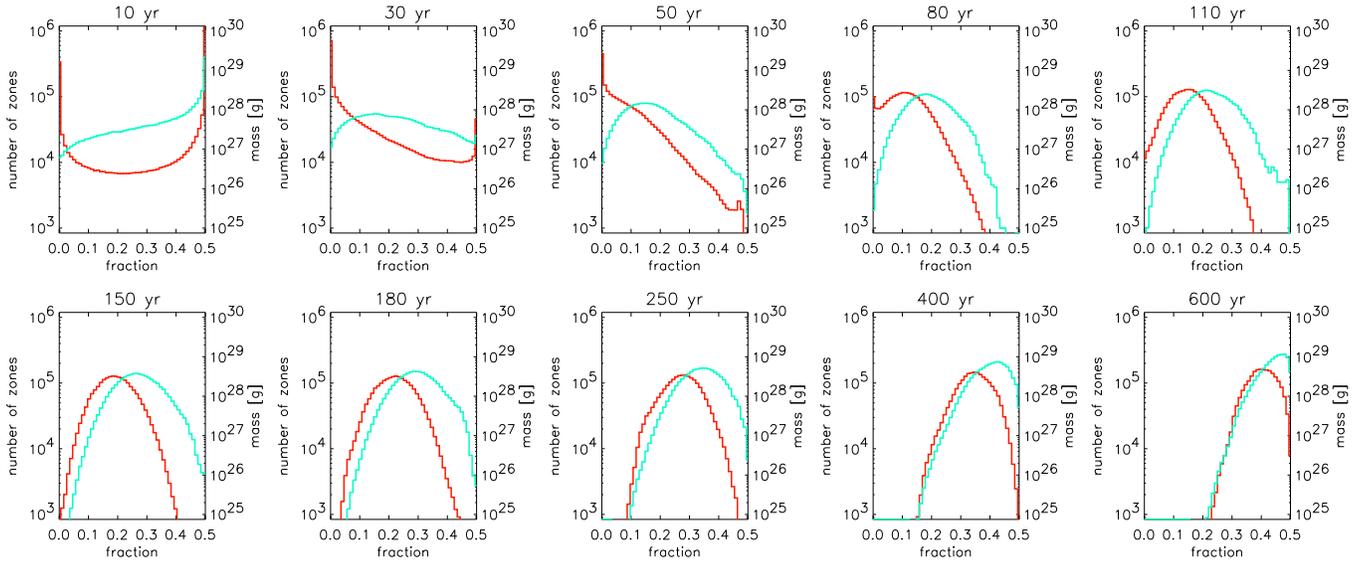}
\end{raggedleft}
\caption{Histograms of the molecular fraction distribution.
The black curve corresponds to the left y-axis and represents the volume
distribution of molecules: the number of zones with 
value of molecular fraction within the interval $[f, f + \Delta f]$, 
$\Delta f = 10^{-2}$. The grey curve corresponds to the right
y-axis and represents the mass distribution of molecules.
The data were taken from the $128^3$ run with 
rms velocity of 60 km~s$^{-1}$.}
\label{fig:histmf}
\end{figure*}
\begin{figure}
\centering\psfig{file=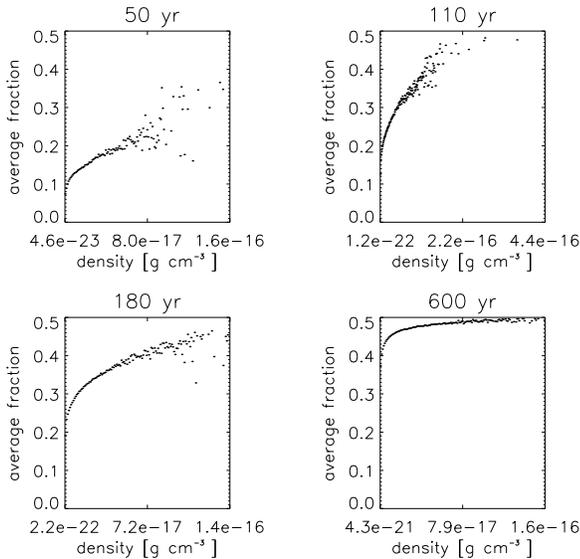}
\caption{Histograms of the average molecular fraction distribution
with density at different times. Molecules reform quickly even
in regions with low densities. Data from 128$^3$ resolution run with
initial rms velocity of 60 km~s$^{-1}$.}
\label{fig:histdf}
\end{figure}
%
%-------------------------------------------------------------------
The molecular fraction `$f\,$' is displayed in Fig.~\ref{fig:frac}.
The three initial states correspond to three distinct physical
regimes. With a  rms velocity of 15 km~s$^{-1}$, very few dissociative
shocks develop but some localised dissociation still occurs.
With 30 km~s$^{-1}$, a few per cent of the molecules are dissociated
whereas at 60 km~s$^{-1}$, the gas becomes over 80\% atomic.

Note that the minimum value of $f$ is reached after $\sim 4\,t_i$.
The dynamics alone determines the period of the dissociation phase,
which does not overlap with the reformation phase for the chosen
parameters.

Reformation of molecular hydrogen is unexpectedly rapid.
The expected H$_2$ reformation time at 20\,K and 10$^6$\,cm$^{-3}$ is
$t_R = (k_R\,n_o)^{-1}$ = 3,200\,yr (Eq.\,1) a factor of 5 larger than
the  simulation time.
At 100\,yr, the temperature is $\sim 80$\,K, predicting a
reformation time of $t_R = (k_R\,n_o)^{-1}$ = 2,000\,yr.  Yet,
reformation is occurring over $\sim 400$~yr. This speed up
is caused by the turbulence itself: the molecules preferentially
reform in the denser and cooler locations. As weak shocks
propagate through the gas, different regions are compressed
and expanded. Hence the reformation time is not only
controlled by the `average' reformation time, but also by the
strength of the turbulence. Given a turbulent dynamical timescale
shorter than the average reformation timescale, then we can
expect reformation to be accelerated.

We have checked the degree of convergence of both the
dissociation and reformation processes. Molecule reformation is a
gradual process and is, therefore, accurately represented in the simulations.
High resolution is, however, paramount to correctly
follow the degree of dissociation. This is critical to
the 30\,km\,s$^{-1}$ turbulence since there are numerous intermediate
speed shocks, partially dissociating the gas. For the low and high
speed examples, however, the dissociation is basically zero and
complete, respectively. Exhaustive resolution studies of
one-dimensional shocks with this code
are presented elsewhere \citep{Smith02}.

The remarkable manner in which the molecular fraction changes is
illustrated in Fig.~\ref{fig:histmf} for the high speed case.
This figure displays the number of zones $N_z(f)$
with a molecular fraction in each interval
$[f,\,f\!+\Delta f]$ where $\Delta f$ is taken as 10$^{-2}$.
After 10\,yr, there is an even distribution, suggesting that
dissociative  shocks have influenced about half the gas.
Up to 50 years, the majority of zones have low values of $f$.
Thereafter, the reformation has the following  properties.
\begin{itemize}
\item A single maximum in $N_z(f)$ shifts with time to higher
values of $f$.
\item The maximum remains at $N_z\sim 10^5$ zones.
\item The width of the peak is roughly constant (full
width at 80\% maximum is 0.16\,$\pm$\,0.02)
\item The distribution of $f$ by mass displays a single wide peak
(grey line in Fig.~\ref{fig:histmf}).

\item Even in regions with a very low density, reformed molecules
are found after 100~yr (Fig.~\ref{fig:histdf}).
\end{itemize}
These vital results imply that $f$ is dominated by a small
range in values at any instant, at times after 100~yr. 
For this reason, three
dimensional structure and maps
based on column density of either all the gas or just the
molecules appear almost identical.
In this sense, isothermal simulations can be utilised to
determine the {\em structure} of molecular clouds or cores
but not the underlying fraction of molecules.

%-----------------------------------------------------------------
\subsection{Statistical analysis}\label{ssec:cf}
%
%===========================================
\begin{figure*}
\centering\psfig{file=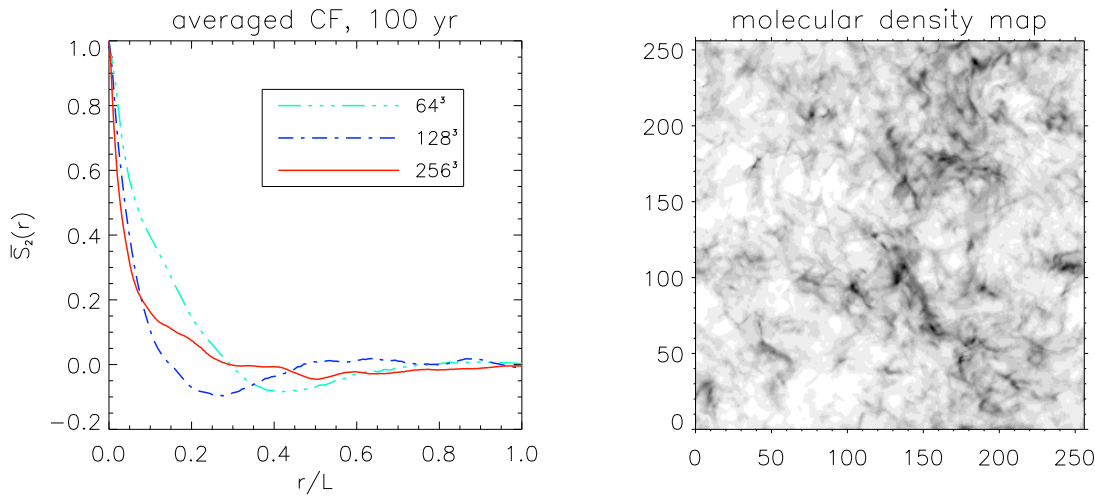}
\caption{Autocorrelation functions for molecular density and snapshot
of molecular density map (from 256$^3$ data) at 100 yr. 
The molecular density map is the result of the 
integration of the data cube (fraction $\times$
density) along $z$ axis.}
\label{fig:cf100} %Fig 6
\end{figure*}
%
%===========================================
\begin{figure*}
\centering\psfig{file=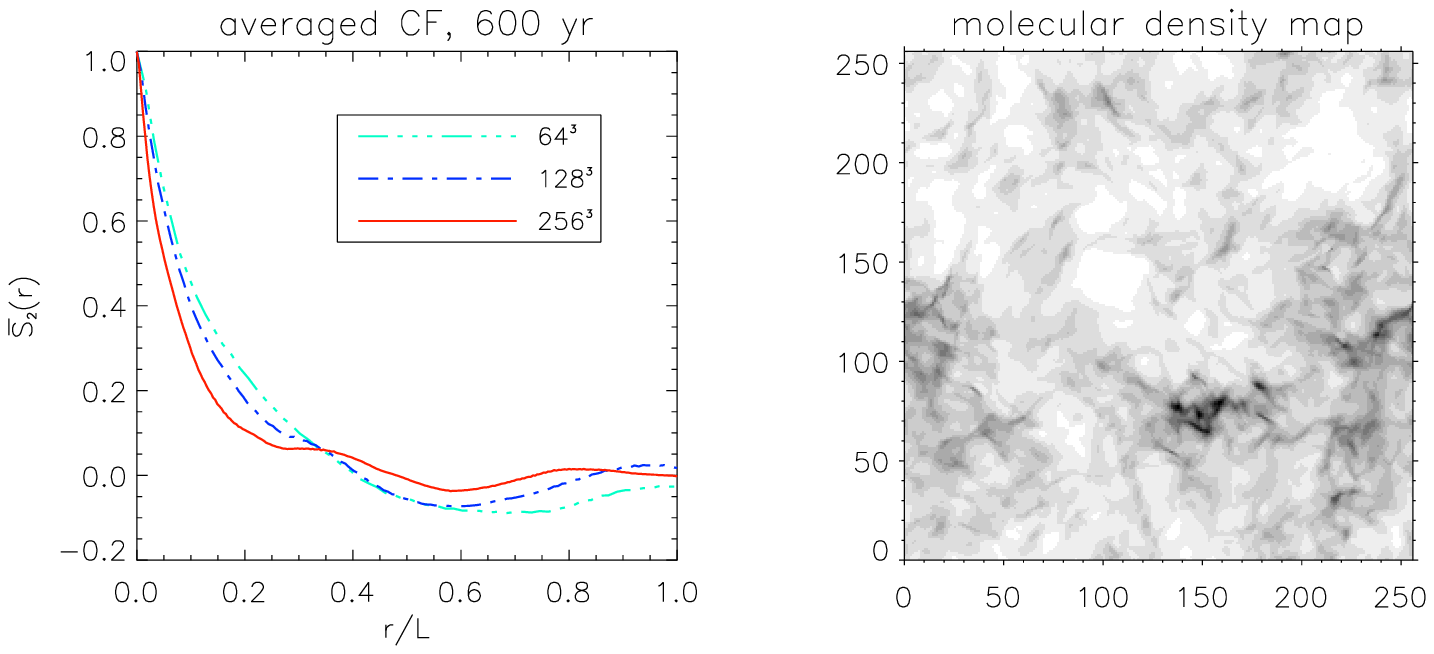}
\caption{Autocorrelation functions for the molecular 
density and snapshot of the molecular density map 
(from 256$^3$ data) at 600 yr. The molecular 
density map is the result of the  integration of the data cube 
(fraction $\times$ density) along $z$ axis.}
\label{fig:cf600} %Fig 7
\end{figure*}
%
%===========================================
\begin{figure}
\centering\psfig{file=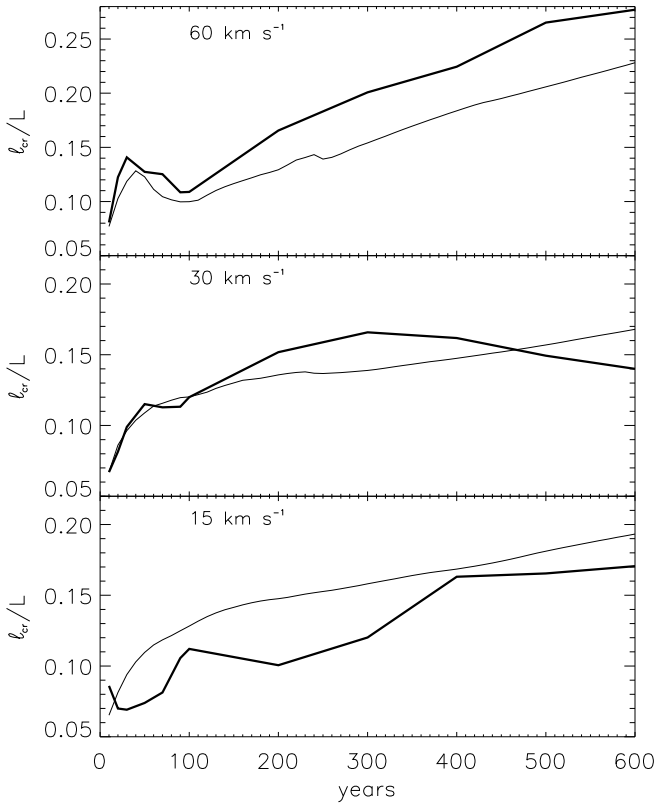}
\caption{The correlation length changes with time in runs with different
initial rms velocity. The data taken from the 128$^3$ run. Thick lines 
represent correlation length, thin line represent inverse average 
Mach number change.}
\label{fig:lcr} %Fig 8
\end{figure}
%
%===========================================
\begin{figure*}
\begin{minipage}[t]{0.5\linewidth}
\centering\psfig{file=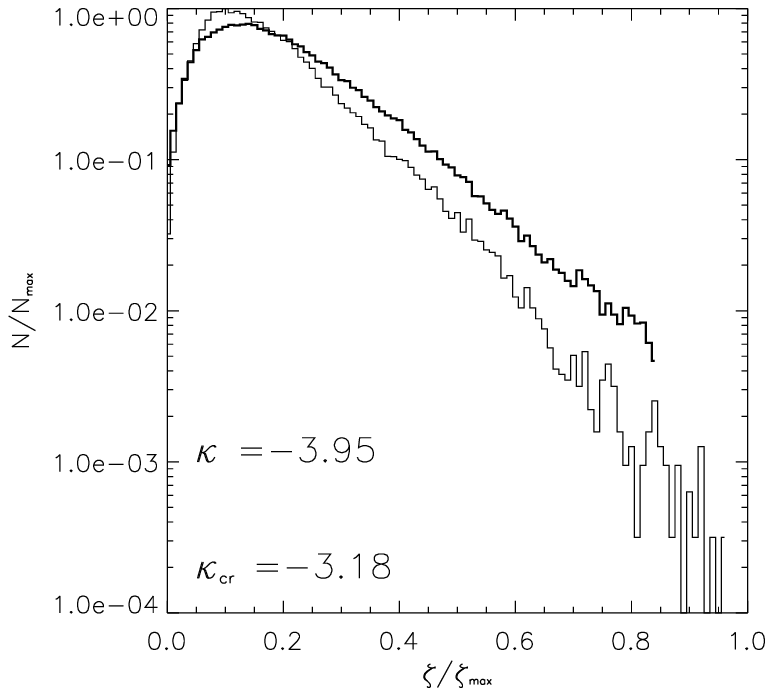}
\end{minipage}\hfill
\begin{minipage}[t]{0.5\linewidth}
\centering\psfig{file=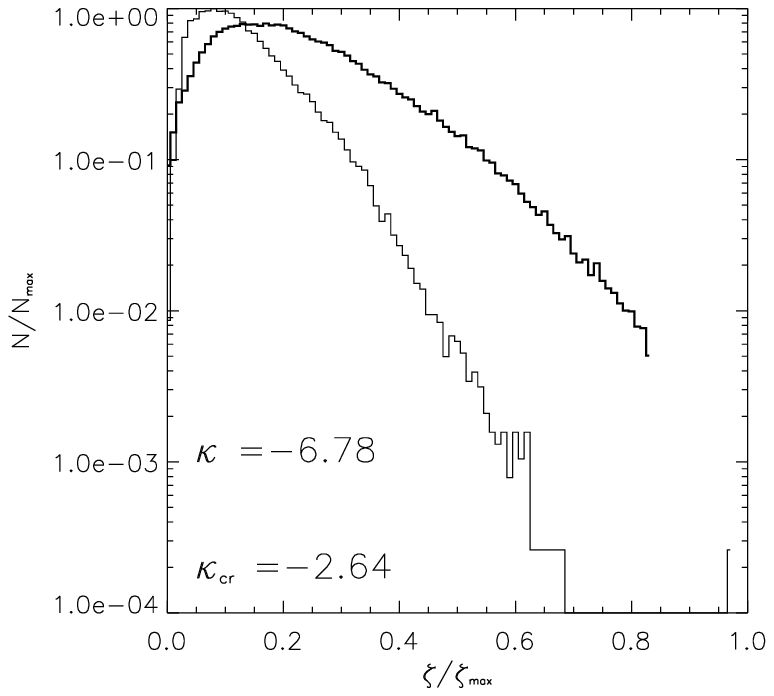}
\end{minipage}
\caption{Probability distribution functions. The left figure presents
histograms from the 60~km~s$^{-1}$ run at 100\,yr. 
The right figure presents
histograms from the 30~km~s$^{-1}$ run at 100\,yr. The thicker line 
corresponds to the averaged PDF from areas with a size of
about or less then one correlation length ($ \approx 0.23 L$),
with the associated linear fit coefficient $\kappa_{cr}$.
The data are taken from the 256$^3$ run.}
\label{fig:pdf} %Fig. 9
\end{figure*}
%===========================================
%
In recent years, several studies of probability density functions
in numerical simulations have been advanced as a step in their
full statistical characterisation (e.g. \cite{VazSem01, Burkert01}).
The goal is to quantitatively compare observed maps of the 
interstellar medium with the predictions of simulations. 
Here, we present the first analysis of the results for the 
column density of molecules from turbulent simulations which
include the molecular chemistry.

The autocorrelation function is defined as follows:
\begin{equation}\label{equ:cf}
S_2(x, y) = \frac{\displaystyle\sum_{i,j}^{L}
\big(\zeta(i, j) - \langle\zeta\rangle\big)
\big(\zeta(i + x, j + y) - \langle\zeta\rangle\big)}
{\displaystyle\sum_{i, j}^{L}
\big(\zeta(i, j) - \langle\zeta\rangle\big)^2},
\end{equation}
where $\zeta(x, y)$ is the quantity (integrated along the LOS):
\begin{equation}\label{equ:zeta}
\zeta(x, y) = \frac{1}{L}\sum_{z}^{L}\xi(x, y, z),
\end{equation}
where $\xi(x, y, z)$ is the density (or molecular density), and
$\langle\zeta\rangle$ is an average:
\begin{equation}\label{equ:avzeta}
\langle\zeta\rangle = \frac{1}{L^2}\sum_{x, y}^{L}\zeta(x,y).
\end{equation}
As a qualitative measure, however, it is more convenient to 
use an averaged correlation function, i.e. polar angle integrated:
\begin{equation}\label{equ:avcf}
\bar{S}_2(r) = \frac{2}{\pi}\int_{0}^{\pi/2} S_2(r\cos{\phi},
r\sin{\phi}) r\,{\rm d }\phi,\quad r\in[0, L].
\end{equation}
Having integrated the correlation function, we have lost all
information about anisotropic properties of the molecular
density map. However, in our studies we have found that 
considerable deviations from polar symmetry of a correlation
function occur usually only in regions of weak correlation, i.e.
separated by a distance which is large in comparison with
the correlation length. This suggests that large scale structure
which might seem to appear in molecular density maps (e.g. see the map
on Fig.~\ref{fig:cf100}) is not coherent, and therefore 
turbulence in the simulations can be considered as isotropic. 

We find that the correlation length for the molecular 
column density  grows as the  turbulence decays  
(see Fig.~\ref{fig:lcr} also  compare Fig.~\ref{fig:cf100}
and Fig.~\ref{fig:cf600}) as has been predicted for gas in general
by \cite{M-MML99} and \cite{M-MML00}. Defining the correlation length 
as the length over which the  average autocorrelation function 
has decayed by a factor of 10 ($\bar{S}_2(\ell_{cr}) = 0.1$),
we find that in the run with resolution of 128$^3$,
the correlation length is related to the
average sound speed.
The growth of $\ell_{cr}$  shows a good correlation with the
inverse average Mach number behaviour (see Fig.~\ref{fig:lcr}). An
explanation for this is easily seen from the following analysis.
Defining the crossing time as $t_{cross}= L/\bar{v}$,
where $\bar{v}$ is the average speed in the box, we find that the
`information' during this time has travelled as much
as $\ell_{cr} = \bar{c}_s \times t_{cross}$.
From this, one immediately obtains the following relation:
\begin{equation}\label{equ:lcr}
\frac{\ell_{cr}}{L} = \frac{\bar{c}_s}{\bar{v}}\approx M\!a ^{-1}.
\end{equation}

The autocorrelation functions for density and molecular density
maps indeed look very similar, as expected given the above results
for the evolution of the molecular fraction within a relatively
narrow range of values. \\

An analysis of the probability density functions (PDF) reveals 
no particular
difference between PDFs for density and molecular density maps and
PDFs from different moments of time.
Linear fit coefficients $\kappa$ of the histograms 
of $\zeta$ in log-linear coordinates 
($N\propto(\zeta / \zeta_{max})^{\kappa} $) are distributed in 
a quite wide range: $\kappa\approx[4.0, 6.8]$ with a mean 
value of 5.13 and dispersion $\approx0.66$ (see Fig.\,\ref{fig:pdf}).

It has been proposed that observational measurements are
often confined to within a correlation length
(\cite{Burkert01}).
Linear fit coefficients of the average histograms (mean from
a hundred histograms) of regions which have size of about a correlation
length, yield a distribution with a  smaller dispersion 
($\approx0.22$)  and have a mean value of 2.90 
(Fig.\,\ref{fig:pdf}). This is  roughly  consistent with 
the observed value for molecular clouds found by 
\cite{Will00, Blitz97}. In their studies they have used two-dimensional
column density maps of clouds observed in different radial velocity
bins for an optically thin molecular species. Each pixel in the 
data cube (galactic latitude, longitude, radial velocity) has an 
associated antenna temperature $T_a$, proportional to the column
density of gas in that pixel. The PDFs was then defined as the
total number $N$ of pixels with a certain column density
$\zeta / \zeta_{max}$, (where $\zeta_{max}$ 
is the maximum column density
found in the data cube). For molecular clouds in Taurus the following
asymptotic\footnote{$\kappa$ is found to be sensitive to the spatial
smoothing, for smoothing beam $\gtrsim 20'$ the PDF reaches an
asymptotic value}
value of $\kappa$ is reached: $\kappa = 2.7\pm 0.4$. Although our
simulations are not directly comparable with the observations of Giant
Molecular Clouds, the consistency is quite remarkable.

%-----------------------------------------------------------------
\subsection{Spatial Distribution}
\begin{figure} % Fig. 10
\centering\psfig{file=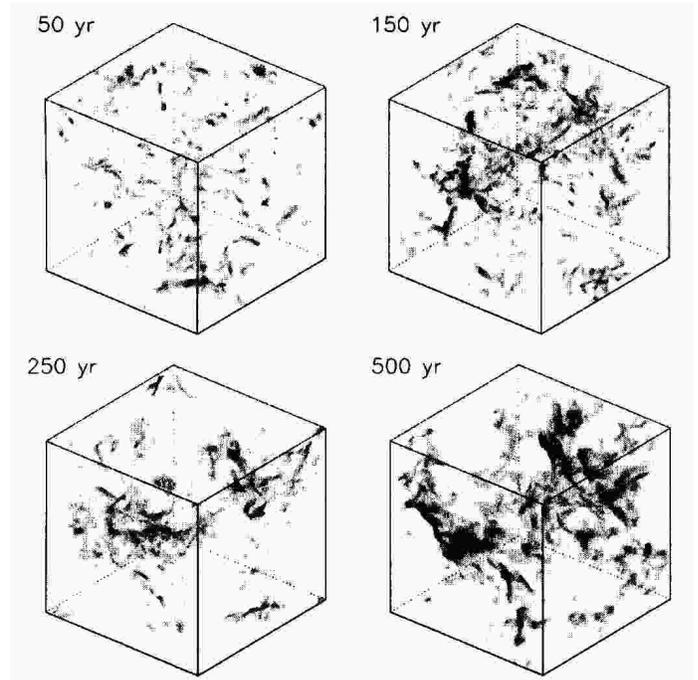}
\caption{Molecular density distribution snapshots. Data taken from
the 128$^3$ resolution run with rms velocity 60 km~s$^{-1}$. At early
stages (when average molecular fraction is low) the distribution of
molecular density is different from mass density (see 
Fig.~\ref{fig:den}); later, when hydrogen reforms, the distributions 
look very similar}
\label{fig:molden}
\end{figure}

\begin{figure} %Fig. 11
\centering\psfig{file=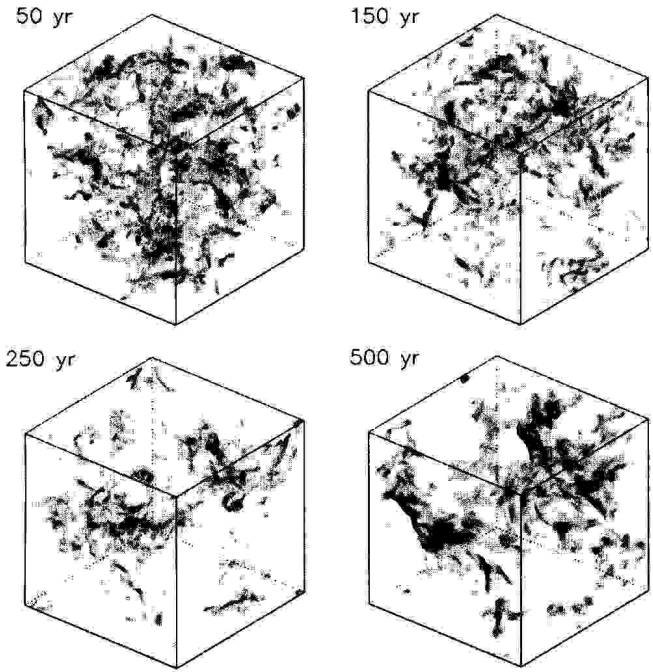}
\caption{Mass density distribution snapshots. Data taken from
128$^3$ resolution run with rms velocity 60 km~s$^{-1}$. Compare the
distributions with the molecular density distribution on 
Fig.~\ref{fig:molden}.}
\label{fig:den}
\end{figure}

In Figs. \ref{fig:molden} and  \ref{fig:den}, we present 
representations of the three dimensional distributions.
These figures demonstrate how the molecular density
and total density distributions converge between 50 and 100\,yr. 
Initially, of course, while thermal  dissociation is occurring
in the shock waves, there is less correlation.

The general structure of the clouds is hard to define.
There are some clumps and filament-like structures but little visual
evidence for sheets. Some clumps and filaments are surrounded 
by diffuse envelopes, which are adjacent to other filaments 
and clumps. Self-gravity or sweeping by large-scale shocks could
generate more coherent structures. 

%
%===========================================
%
%===========================================
%
%
%===========================================
\section{Conclusions}

\subsection{Summary}

We  have presented the properties of
a specific model for molecular turbulence.
We carried out three dimensional hydrodynamical
simulations of decaying supersonic turbulence in
molecular gas. We included a detailed cooling
function, molecular hydrogen chemistry and equilibrium C and O
chemistry. We studied three cases in which the applied
velocity field straddles the value for which
wholesale dissociation of molecules occurs.
The parameters chosen ensure that for the
high-speed turbulence,
the molecules are initially destroyed in shocks and gradually reform
in a distinct phase.

We find the following.
\begin{itemize}
\item An extended phase of power-law kinetic energy
decay, as in the isothermal case, after an initial phase of slow
dissipation and shock formation.
\item  The thermal energy, initially raised by
the introduction of turbulence, decays only a little slower
than the kinetic energy.
\item The reformation of hydrogen molecules, as the fast turbulence
decays, is several times faster than expected from
the average density. This is expected in a non-uniform medium,
as a consequence of the volume
reformation rate being proportional to the density squared.
\item The molecules reform into a
pattern of  filaments and small clumps, enveloped in
diffuse structure.
\item  During reformation, the remaining turbulence redistributes
the gas so that the fraction of molecules is
distributed relatively evenly. Hence, the density and 
molecular density are almost identically distributed at 
any one time after 100 yr.
\item The correlation length in our simulations grows 
with time as has been  predicted for a gas in general.
\item The probability density functions sampled from regions with size
of about one correlation length are consistent with observations.
\end{itemize}
%
%-----------------------------------------------------------------
\subsection{Discussion}
We mainly wish here to emphasise the insight these simulations provide
into how molecular chemistry and supersonic dynamics combine.
We have found that isothermal simulations are indeed very useful,
not only for the rate of energy decay but also to trace the
molecules.

A simple reason for the fast decay is that a sufficient number of
strong shocks survive. As shown by \cite{Smith00I}, the
rate of energy decay in {\em decaying} turbulence is dominated by
the vast number of weak shocks. These shocks are less efficient
at energy dissipation. A second possible reason is that the curved
shock structures create small scale vorticity, which leads to enhanced
dissipation of kinetic energy. It is not clear, however, why relatively
more vorticity should be created when stronger shocks are present.
Simulations of isolated curved shocks would help clarify the
dissipation paths.

It is plausible that the high Mach number turbulence could create
more small-scale structure, leading to a faster decay rates as argued
by \cite{M-MML99}.

The simulations presented here may directly aid our interpretation
of fast molecular shocks in dense regions, such as Herbig-Haro objects.
For example, supersonic decay would appear 
to occur in the wake of bow shocks
such as in DR~21 \citep{Davis96}. These simulations 
suggest that the turbulence
decays rapidly and bow shock wakes will be bright but short.

The fast reformation of molecules indicates that molecules may form on
shorter time scales than previously envisaged. Hence molecular clouds
may appear out of atomic clouds several times faster than anticipated
in non-turbulent scenarios. Recently, some evidence has been found
for rapid cloud formation and dissipation \citep{Balles99a, Hart01}.
This implies that much of the interstellar medium
may be undergoing both rapid dynamical and chemical changes, driven by
sources of supersonic turbulence.

The simulations analysed here provide a basis for much 
further work. For example,
we can now investigate the properties of 
hydrodynamic clouds in which dust or
chemical abundances are non-uniform, clouds in which 
the turbulence is driven
uniformly or non-uniformly, and clouds which are initially atomic.
%
%-----------------------------------------------------------------
\section{Acknowledgments}
The computations reported here were performed using the UK
Astrophysical Fluids Facility (UKAFF) and FORGE (Armagh),
funded by the PPARC JREI scheme, in collaboration with SGI. M-MML
was partially funded by the NASA Astrophysical Theory Program
under grant number NAG5-10103 and NFS CAREER grant AST99-85392.
AR was funded by the PPARC. Armagh Observatory receives funding
from the Northern Ireland Department of Culture, Arts and Leisure.
 ZEUS-3D was used by courtesy of the
Laboratory of Computational Astrophysics at the NCSA.
%
%----------------------------------- should be processed by bibtex
%
\bibliography{gbp}
\appendix
\section{The Chemisitry and The Cooling Function}\label{appx}
We consider only a limited network of chemical reactions, which has been
tested in one-dimensional simulations \citep{Smith02} and which is 
critical for molecular cloud evolution.  These reactions are:
\begin{align}
\label{f:ch.h1}
\rmn{H} + \rmn{H} &\longmapsto \rmn{H}_2 \\ 
\label{f:ch.h2}
\rmn{H}_2 + \rmn{H} &\longmapsto 3\rmn{H} \\
\label{f:ch.h3}
\rmn{H}_2 + \rmn{H}_2 &\longmapsto 2\rmn{H} + \rmn{H}_2 \\
\label{f:ch.o}
\rmn{O} + \rmn{H}_2 &\longmapsto \rmn{OH} + \rmn{H} \\
\label{f:ch.oh1}
\rmn{OH} + \rmn{H} &\longmapsto \rmn{O} + \rmn{H}_2 \\
\label{f:ch.oh2}
\rmn{OH} + \rmn{C} &\longmapsto \rmn{CO} + \rmn{C} \\
\label{f:ch.co}
\rmn{CO} + \rmn{H} &\longmapsto \rmn{OH} + \rmn{C} \\
\label{f:ch.h2o1}
\rmn{OH} + \rmn{H}_2 &\longmapsto \rmn{H}_2\rmn{0} + \rmn{H} \\
\label{f:ch.h2o2}
\rmn{H}_2\rmn{O} + \rmn{H} &\longmapsto \rmn{OH} + \rmn{H}_2
\end{align}
The first three reactions (\ref{f:ch.h1}, \ref{f:ch.h2}, \ref{f:ch.h3})
are included in a semi-implicit cooling-chemistry step to calculate 
the temperature and H$_2$ fraction. Reaction (\ref{f:ch.h1}) takes 
place on  grain surfaces with the rate given by equation (\ref{eqn:kR})
(see section~\ref{ssec:zeus}).

We fix free oxygen and carbon abundances, $\eta_0(\rmn{O})$ and 
$\eta_0(\rmn{C})$. We then calculate the equlibrium O, OH, and CO
abundances, assuming that the CO reactions are much faster than the
H$_2$O reactions. Then, the remaning free oxygen is distributed into
O, OH, and H$_2$O, according to equlibrium abundances. In this manner
we can calculate quite accurately the chemistry within fast shocks
without overloading the hydrocode. For further details 
\citep[see][Appendix~B]{Smith02}.

The cooling fuction used in the numerical code (see equations 
\ref{equ:hydro1} --  \ref{equ:hydro4} ) is composed of 
11 separate parts (one of which heats the gas):
\begin{equation}\label{equ:coolf}
\Lambda = \sum_{i=1}^{11}\Lambda_i
\end{equation}
The components are summarised in Table~\ref{tab:coolf} below.
\begin{table*}
\begin{minipage}{170mm}
\caption{Components of the cooling function}
\label{tab:coolf}
\begin{tabular}{p{9cm} p{7.7cm}}
    {\bf Formulae} %
    & {\bf Description} \\
\hline
%%----------------------------------------------------------------------
    $\Lambda_1 = \lambda_1 \times n^2$, where, on assuming 
      standard dust properties,
      \[
      \lambda_1 = 3.8\times10^{-33.0}T^{0.5}
      (T - T_{\rmn{dust}}(1.0 - 0.8\exp(-75 / T))
      \,\,[\rmn{erg\,s}^{-1}]
      \]
    & {\bf gas-grain (dust) cooling} \citep{Holl89}. 
       In present calculations we  fix $T_\rmn{dust} = 20$K. \\
%%----------------------------------------------------------------------
    $\Lambda_2 = n_{\rmn{H}_2}\left(
                \frac{L_v^{(h)}}{1+L_v^{(h)}/L_v^{(l)}} %
                + \frac{L_r^{(h)}}{1+L_r^{(h)}/L_r^{(l)}}\right)$, 
    where the {\it v}ibrational and {\it r}otational 
    coefficients at {\it h}igh and {\it l}ow density are:
    \[
    L_v^{(h)} = 1.10\times10^{-18}\exp(-6744/T)\,\,[\rmn{erg\,s}^{-1}],
    \]
    \[
    L_v^{(l)} = 8.18\times10^{-13}\exp(6840/T)
               \left(n_{\rmn{H}}k_{\rmn{H}(0,1)} +
                n_{\rmn{H}_2}k_{\rmn{H}_2(0,1)}\right)
               \,[\rmn{erg\,s}^{-1}],
    \] 
   the terms $k_{\rmn{H}(0,1)}$ and $k_{\rmn{H}_2(0,1)}$  are 
   $v:0\rightarrow1$ collisional excitation rates,
    \[
k_{\rmn{H}(0,1)} = 
                 \begin{cases}
                  1.4\times10^{-13}\exp(T/125 - T^2/577^2), 
                  & \text{if $T<T_v$}, \\
                  1.0\times10^{-12}T^{0.5}\exp(-1000/T), 
                  & \text{if $T>T_v$},
                 \end{cases}
    \] 
    where $T_v$ = 1635~K,
    \[
k_{\rmn{H}_2(0,1)} = 1.45\times10^{-12}T^{0.5}\exp(-28728/(T+1190));
    \]
    \[
L_r^{(h)} = 
           \begin{cases}
           \rmn{dex}\{-19.24+0.474x -1.247x^2\},
           & \text{if $T<T_r$ }, \\
           3.9\times10^{-19}\exp(-6118/T),
           & \text{if $T>T_r$ },
           \end{cases}
    \] 
    where $T_r$ = 1087~K, and $x=\log(T/10000)$,
    \[
\frac{L_r^{(l)}}{Q(n)} = 
                       \begin{cases}
                       \rmn{dex}\{-22.90-0.553-1.148x^2\},
                       & \text{if $T<T_l$}, \\
                       1.38\times10^{-22}\exp(-9243/T),
                       & \text{if $T>T_l$},
                       \end{cases}
     \] 
    where $T_l$ = 4031~K, and $Q(n) = \left(n_{\rmn{H}_2}\right)^{0.77}
     + 1.2\left(n_{\rmn{H}}\right)^{0.77} $ 
    & {\bf collisional cooling associated with vibrational
       and rotational modes of molecular hydrogen}. These formulae
       based on equations 7 -- 12 in \cite{Lepp83} \\
%%----------------------------------------------------------------------
    $\Lambda_3 = \left(n_{\rmn{H}}\right)^2\times \lambda_2$
    & {\bf collisional cooling of atoms}. We have used Table~10
      of \cite{Suth93} (with $\rmn{Fe} = -0.5$) for the form of 
      $\lambda_2$ and we added an extra thermal bremsstrahlung
      term equal to $1.42\times10^{-27}T^{0.5}$ for $T>10^4$~K \\
%%----------------------------------------------------------------------
    $\Lambda_4 = (n_{\rmn{H}_2} + 1.39 n_{\rmn{H}})\times
                    n_{\rmn{H}_2\rmn{O}}\times\lambda_3$;
            $\lambda_3 = 1.32\times10^{-23}(T/1000)^\alpha$ ,
      where $\alpha = 1.35 - \log(T/1000)$.
    & {\bf cooling through rotational modes of water induced
      by collisions with both atomic and molecular hydrogen}.
      Given values of $\alpha$ fits the values tabulated by
      \cite{Neuf93}\\
%%----------------------------------------------------------------------
    $\Lambda_5 = 1.03\times10^{-26}n_{\rmn{H}_2}
               n_{\rmn{H}_2\rmn{O}}T\exp(-2325/T)\exp(-47.5/T^{1/3})$
    & {\bf cooling through vibrational modes of water induced
      by collisions with molecular hydrogen} \citep{Holl89} \\
%%----------------------------------------------------------------------
    $\Lambda_6 = 7.40\times10^{-27}n_{\rmn{H}}\lambda_3
               n_{\rmn{H}_2\rmn{O}}T\exp(-2325/T)\exp(-34.5/T^{1/3})$
    & {\bf cooling through vibrational modes of water induced
      by collisions with atomic hydrogen} \citep{Holl89} \\
%%----------------------------------------------------------------------
     $\Lambda_7 = 7.18\times10^{-12}\left((n_{\rmn{H}_2})^2 
                   k_{D,\rmn{H}_2}+n_{\rmn{H}}n_{\rmn{H}_2}
                   k_{D,\rmn{H}}\right),$
      where dissociation coefficients are taken to be,
     \[
k_{D,\rmn{H}} = 1.2\times10^{-9}\exp(-52400/T)
                \left(0.0933\exp(-17950/T)\right)^\beta\,\,
                [\rmn{cm}^3\,\rmn{s}^{-1}],
     \]
     \[
k_{D,\rmn{H}_2} = 1.3\times10^{-9}\exp(-53300/T)
                  \left(0.0908\exp(-16200/T)\right)^\beta\,\,
                 [\rmn{cm}^3\,\rmn{s}^{-1}],
     \]
     \[
\beta = \left[1.0+n\left(2f\left(n_2^{-1}-n_1^{-1}\right)+
              n_1^{-1}\right)\right]^{-1}
     \]
     critical densities are fit by the following approximation,
     $n_1 = \rmn{dex}\{4.0-0.416x-0.327x^2\}\,\,[\rmn{cm}^{-3}]$ and
     $n_2 = \rmn{dex}\{4.845-1.3x+1.62x^2\}\,\,[\rmn{cm}^{-3}]$.
    & {\bf cooling from the dissociation of molecular hydrogen}. Factor
     $7.18\times10^{-12}\,[\rmn{erg}]$ is the 4.18 eV dissociation
     energy; $n_1$ --  is the density critical for dissociation by 
     collision of molecular hydrogen with atomic hydrogen, $n_2$ -- 
     with itself.
      \\
%%----------------------------------------------------------------------
    $\Lambda_8 = -\lambda_7\xi$, where $\lambda_7=k_R$ 
      (see Eq.~\ref{eqn:kR}) and $\xi = nn_{\rmn{H}}(1-\beta)
      7.18\times10^{-12}$
    & {\bf heating resulting from the reformation of molecular
       hydrogen}. We employ $\beta$ to parametrise the fraction of 
       released energy which is thermalised rather than radiated.
      \\
%%----------------------------------------------------------------------
     $\Lambda_9 = n_{\rmn{CO}}nkT\sigma v_T/
                \left(1+n_a/n_{cr}+1.5(n_a/n_{cr})^{0.5}\right)$, where
      the mean speed of the molecules $v_T = \sqrt{8kT/(\pi
      m_{\rmn{H}_2})}$ and $n_{cr} = 3.3\times10^{6}(T/1000)^{0.75}
      \,[\rmn{cm}^{-2}]$, $\sigma=3.0\times10^{-16}(T/1000)^{-0.25}
      \,[\rmn{cm}^{-2}]$. Average density, $n_a = 0.5(n_{\rmn{H}}+
      n_{\rmn{H}_2}\sqrt{2})$.
    & {\bf cooling through rotational modes of carbon monoxide induced
       by collision with both atomic and molecular hydrogen}. We based
       our equations on Eqs.~5.2--5.3 in \cite{McKee82} \\
%%----------------------------------------------------------------------
     $\Lambda_{10} = 1.83\times10^{-26}n_{\rmn{H}_2}n_{\rmn{CO}}T
      \exp(-3080/T)\exp(-68/T^{1/3})$
    & {\bf cooling through vibrational modes of carbon monoxide induced
       by collisions with molecular hydrogen} \citep[see][]{Neuf93} \\
%%----------------------------------------------------------------------
     $\Lambda_{11} = 1.28\times10^{-24}n_{\rmn{H}}n_{\rmn{CO}}T^{0.5}
      \exp(-3080/T)\exp(-(2000/T)^{3.43})$
    & {\bf cooling through vibrational modes of carbon monoxide induced
       by collisions with atomic hydrogen} \\
%%----------------------------------------------------------------------
%\hline
\end{tabular}
\end{minipage}
\end{table*}

\end{document}